\documentclass{article}
\usepackage{amsfonts}
\usepackage{makeidx}
\usepackage{latexsym,amsmath,amssymb,amscd}
\usepackage{makecell}
\usepackage{xcolor}
\usepackage{multirow}
\usepackage[all]{xy}

\allowdisplaybreaks
\newtheorem{theorem}{Theorem}

\begin{document}

\title{The generalized Ermakov conservative system: A discussion}
\author{Antonios Mitsopoulos$^{1,a)}$ and Michael Tsamparlis$^{1,b)}$  \\
{\ \ }\\
$^{1}${\textit{Faculty of Physics, Department of
Astronomy-Astrophysics-Mechanics,}}\\
{\ \textit{University of Athens, Panepistemiopolis, Athens 157 83, Greece}}
\vspace{12pt} 
\\
$^{a)}$Email: antmits@phys.uoa.gr 
\\
$^{b)}$Email: mtsampa@phys.uoa.gr }
\date{}
\maketitle

\begin{abstract}
Using older and recent results on the integrability of two-dimensional (2d) dynamical systems, we prove that the results obtained in a recent publication concerning the 2d generalized Ermakov system can be obtained as special cases of a more general approach. This approach is geometric and can be used to study efficiently similar dynamical systems.
\end{abstract}

\section{Introduction}

The two-dimensional (2d) generalized Ermakov system has attained attention in 90's (\cite{Ermakov}, \cite{Ray 1979A}, \cite{Athorne 1990}, \cite{Leach 1991}, \cite{Leach 1994}, \cite{Goedert 1998}) where most of its properties have been revealed. A review of these studies can be found in \cite{Leach Andriopoulos 2008}. However, the interest in the topic is still alive and a recent article has appeared in this journal \cite{Naz 2020} presenting new results. The purpose of the present discussion is to show that these latter results can be obtained as special cases of older and more recent results on the integrability of 2d dynamical systems.

\section{The 2d generalized Ermakov system}

\label{sec.erma.automous}

The 2d generalized Ermakov system is defined by the equations
\begin{eqnarray}
\ddot{x} &=& -\omega^{2}(t)x + \frac{1}{x^{2}y} f\left(\frac{y}{x}\right)
\label{eq.erm1a} \\
\ddot{y} &=& -\omega^{2}(t)y + \frac{1}{xy^{2}} g\left(\frac{x}{y}\right)
\label{eq.erm1b}
\end{eqnarray}
where $f, g$ are arbitrary functions. This system admits the Ermakov first
integral (FI)
\begin{equation}
I_{0}= \frac{1}{2} (x\dot{y}-y\dot{x})^{2} + \int^{y/x} f(u)du + \int^{x/y}
g(v)dv  \label{eq.erm2}
\end{equation}
where $u= v^{-1}= \frac{y}{x}$. It is well-known that the 2d generalized Ermakov system generalizes the one-dimensional (1d)
time-dependent oscillator.

Introducing the functions $F, G$ by the relations
\begin{equation*}
f\left(\frac{y}{x}\right)= \frac{y}{x} F\left(\frac{y}{x}\right), \enskip %
g\left(\frac{x}{y}\right)= \frac{x}{y} G\left(\frac{y}{x}\right)
\end{equation*}
equations (\ref{eq.erm1a}), (\ref{eq.erm1b}) take the
equivalent form
\begin{eqnarray}
\ddot{x} &=& -\omega^{2}(t)x + \frac{1}{x^{3}} F\left(\frac{y}{x}\right)
\label{eq.erm3a} \\
\ddot{y} &=& -\omega^{2}(t)y + \frac{1}{y^{3}} G\left(\frac{y}{x}\right)
\label{eq.erm3b}
\end{eqnarray}
while the Ermakov FI (\ref{eq.erm2}) becomes
\begin{equation}
I_{0}= \frac{1}{2} (x\dot{y}-y\dot{x})^{2} + \int^{y/x} \left[ uF(u) -
u^{-3}G(u)\right]du.  \label{eq.erm4}
\end{equation}

If one introduces the variables \cite{Leach 1991}
\begin{equation}
T= \int \rho^{-2}dt, \enskip X=\rho^{-1}x, \enskip Y=\rho^{-1}y
\label{eq.erm5}
\end{equation}
where $\rho(t)$ is a solution of the 1d time-dependent oscillator
\begin{equation}
\ddot{\rho} + \omega^{2}(t)\rho =0  \label{eq.erm6}
\end{equation}
then equations (\ref{eq.erm3a}), (\ref{eq.erm3b}) become
the autonomous system
\begin{eqnarray}
X^{\prime \prime }&=& \frac{1}{X^{3}} F\left(\frac{Y}{X}\right)
\label{eq.erm7a} \\
Y^{\prime \prime }&=& \frac{1}{Y^{3}} G\left(\frac{Y}{X}\right)
\label{eq.erm7b}
\end{eqnarray}
and the Ermakov FI
\begin{equation}
I_{0}= \frac{1}{2} \left( XY^{\prime} -YX^{\prime} \right)^{2} + \int^{Y/X} \left[ uF(u) -u^{-3}G(u)\right]du \label{eq.erm8}
\end{equation}
where we use the notation $f^{\prime }\equiv\frac{df(T)}{dT}$ and $\dot{f}\equiv
\frac{df(t)}{dt}$. For general functions $F, G$ the autonomous dynamical system (\ref{eq.erm7a}) - (\ref{eq.erm7b}) is not conservative. In the following we determine the
family of conservative Ermakov systems together with their FIs using collineations of the metric defined by these equations.

\section{Integrability of the 2d generalized Ermakov system}

\label{sec.erma.integrability}

Since the system (\ref{eq.erm7a}) - (\ref{eq.erm7b}) is autonomous, the second FI will be the Hamiltonian $H$. To find $H$ we do not have to do any new calculations, because in \cite{Hietarinta}, \cite{MitsTsam 2020} all the integrable and
superintegrable 2d autonomous conservative systems have been determined. From these results, we find (see e.g. section 8 case (1) in \cite{MitsTsam 2020}) that the 2d integrable potential
\begin{equation}
V_{21}= \frac{F_{1}(u)}{X^{2}+Y^{2}} +F_{2}(X^{2}+Y^{2})  \label{eq.ermint3a}
\end{equation}
where $u=\frac{Y}{X}$ and $F_{1}, F_{2}$ are arbitrary functions of their arguments, admits the
FI
\begin{equation}
I_{11}= \frac{1}{2} \left( XY^{\prime }-YX^{\prime} \right)^{2} +F_{1}(u).
\label{eq.ermint3b}
\end{equation}

If we consider $F_{1}(u)= (u^{2}+1)N(u)$ and $F_{2}=0$, we find that $V_{21}=\frac{N(u)}{X^{2}}$ while the resulting equations are
\begin{eqnarray}
X^{\prime \prime }&=& \frac{2N +u \frac{dN}{du}}{X^{3}}  \label{eq.ermint8a}
\\
Y^{\prime \prime }&=& -\frac{1}{X^{3}} \frac{dN}{du}.  \label{eq.ermint8b}
\end{eqnarray}

Therefore, if we define the functions
\begin{equation}
F(u)= 2N +u\frac{dN}{du}, \enskip G(u)= -u^{3}\frac{dN}{du} \label{eq.ermint4b}
\end{equation}
then equations (\ref{eq.ermint8a}), (\ref{eq.ermint8b}) become the Ermakov
equations (\ref{eq.erm7a}), (\ref{eq.erm7b}) while $I_{0}=I_{11}$.

We conclude that the family of the conservative 2d Ermakov systems is
defined by the potential $V=\frac{N(u)}{X^{2}}$ where $N(u)$ is an arbitrary
function while the Hamiltonian is given by the expression
\begin{equation}
H=\frac{1}{2}\left( X^{\prime 2}+Y^{\prime 2} \right) +\frac{N(u)}{X^{2}}.
\label{eq.ermint4c}
\end{equation}

In the original coordinates the system (\ref{eq.ermint8a}) - (\ref%
{eq.ermint8b}) becomes
\begin{eqnarray}
\ddot{x} &=& -\omega^{2}(t)x +\frac{2N(u) +u\frac{dN}{du}}{x^{3}}
\label{eq.ermint7a} \\
\ddot{y} &=&-\omega^{2}(t)y -\frac{1}{x^{3}}\frac{dN}{du}
\label{eq.ermint7b}
\end{eqnarray}
where $u = \frac{y}{x} =\frac{Y}{X}$.

For $N(u)=\frac{u^{-2}}{2}$ we find respectively
\begin{eqnarray}
\ddot{x} &=&-\omega ^{2}(t)x  \label{eq.ermint9a} \\
\ddot{y} &=&-\omega ^{2}(t)y+\frac{1}{y^{3}}  \label{eq.ermint9b}
\end{eqnarray}%
while the Ermakov FI becomes the well-known Lewis invariant \cite{Lewis 1968}
\begin{equation}
I_{0}=\frac{1}{2}(x\dot{y}-y\dot{x})^{2}+\frac{1}{2}\left( \frac{x}{y}%
\right) ^{2}.  \label{eq.ermint9c}
\end{equation}
Equation (\ref{eq.ermint9a}) is the 1d time-dependent harmonic oscillator and (\ref{eq.ermint9b}) is the auxiliary equation with which one determines the frequency $\omega(t)$ for a given function $y(t)$. These justify the characterization of the Ermakov system as a generalization of the harmonic oscillator.

\section{The FIs of the conservative Ermakov system}

There are two ways to find the FIs of the Ermakov system. One way is to use the
results of \cite{Hietarinta}, \cite{MitsTsam 2020} and read the FIs for the
potential $V_{21}$ given in (\ref{eq.ermint3a}) for $F_{1}(u)=(u^{2}+1)N(u)$, $F_{2}=0$. Here, we shall follow another way which can be useful in many similar problems. We shall use Theorem 2 of \cite{Tsamparlis 2011} where it is stated that the generators of the
Noether point symmetries of autonomous conservative systems are the elements
of the homothetic algebra of the metric defined by the kinetic energy
(kinetic metric). In the Ermakov case, this metric is the Euclidean 2d metric $\gamma_{ab}=diag(1,1)$.

For the convenience of the reader we state Theorem 2 of \cite{Tsamparlis
2011}.

\begin{theorem}
\label{thm.point.Noether.autonomous.systems} Autonomous conservative
dynamical systems of the form
\begin{equation}
\ddot{q}^{a} = - \Gamma^{a}_{bc} \dot{q}^{b} \dot{q}^{c} - V^{,a}(q)
\label{eq.PN1}
\end{equation}
where $\Gamma_{bc}^{a}$ are the Riemannian connection coefficients
determined from the kinetic metric $\gamma_{ab}$ (kinetic energy) and $V(q)$
the potential of the system, admit the following point Noether symmetries.
\bigskip

Case 1. The point Noether symmetry
\begin{equation}
\mathbf{A}_{1}=\partial _{t}, \enskip f_{1}=const\equiv0  \label{eq.PN4a}
\end{equation}%
which produces the Noether FI (Hamiltonian)
\begin{equation}
H=\frac{1}{2}\gamma_{ab}\dot{q}^{a}\dot{q}^{b} + V(q).  \label{eq.PN4b}
\end{equation}

Case 2. The point Noether symmetry
\begin{equation}
\mathbf{A}_{2}=2\psi_{B}t\partial _{t}+B^{a}\partial_{q^{a}}, \enskip %
f_{2}=c_{1}t  \label{eq.PN5a}
\end{equation}
where $c_{1}$ is an arbitrary constant and $B^{a}$ is a KV ($\psi_{B}=0$) or
the HV ($\psi_{B}=1$) such that
\begin{equation}
B_{a}V^{,a}+2\psi_{B}V+c_{1}=0.  \label{eq.PN5b}
\end{equation}
The associated Noether FI is
\begin{equation}
I_{2} = 2\psi_{B}tH -B_{a}\dot{q}^{a} +c_{1}t.  \label{eq.PN5c}
\end{equation}

Case 3. The point Noether symmetry
\begin{equation}
\mathbf{A}_{3}=2\psi \int C(t)dt \partial_{t}
+C(t)\Phi^{,a}\partial_{q^{a}}, \enskip f_{3} =C_{,t}\Phi(q)+ D(t)
\label{eq.PN6a}
\end{equation}
where $\Phi^{,a}(q)$ is a gradient KV ($\psi=0$) or a gradient HV ($\psi=1$)
such that ($c_{2}, c_{3}$ are arbitrary constants)
\begin{equation}
\Phi_{,a}V^{,a}+2\psi V=c_{2}\Phi+c_{3}  \label{eq.PN6b}
\end{equation}
and the functions $C(t)$, $D(t)$ are determined by the relations ($%
C_{,t}\neq 0$)
\begin{equation}
C_{,tt}=-c_{2}C, \enskip D_{,t}= -c_{3} C.  \label{eq.PN6c}
\end{equation}
The associated Noether FI is
\begin{equation}
I_{3}=2\psi H\int C(t) dt- C(t) \Phi_{,a}\dot{q}^{a} +C_{,t}\Phi -c_{3}\int
C(t)dt.  \label{eq.PN6d}
\end{equation}
\end{theorem}

We apply Theorem \ref{thm.point.Noether.autonomous.systems} in
the case of the autonomous integrable Ermakov system (\ref{eq.ermint8a}) - (%
\ref{eq.ermint8b}) which has potential $V=\frac{N(u)}{X^{2}}$, where $u=Y/X$,
and kinetic metric $\gamma_{ab}=diag(1,1)$.

The homothetic algebra of $\gamma_{ab}$ consists of two gradient Killing
vectors (KVs) $\partial_{X}, \partial_{Y}$, one non-gradient KV (rotation) $%
Y\partial_{X} -X\partial_{Y}$ and the gradient homothetic vector (HV) $%
X\partial_{X} +Y\partial_{Y}$.

For each case of Theorem \ref{thm.point.Noether.autonomous.systems} we have
the following.

\subsection{The vector $\partial_{T}$}

Case 1. In this case the point Noether symmetry $\mathbf{A}_{1}=\partial_{T}$%
, $f_{1}=0$ produces the Hamiltonian (as expected)
\begin{equation}
H= \frac{1}{2} \left( X^{\prime 2} + Y^{\prime 2} \right) + V= \frac{1}{2}
\left( X^{\prime 2} + Y^{\prime 2} \right) + \frac{N(u)}{X^{2}}.
\label{eq.PNerm1}
\end{equation}

\subsection{The gradient HV $X\partial_{X} +Y\partial_{Y}$}

\label{sec.erm.HV}

Case 2. Consider the gradient HV $B^{a}=(X,Y)$ with homothety factor $\psi_{B}=1$.

Substituting in condition (\ref{eq.PN5b}) we find that
\begin{equation*}
XV_{,X}+ YV_{,Y} + 2V + c_{1}=0 \implies -\frac{2N + u\frac{dN}{du}}{X^{2}}
+ \frac{u}{X^{2}}\frac{dN}{du} +2\frac{N}{X^{2}}+c_{1} =0 \implies c_{1}=0.
\end{equation*}
Therefore, the point Noether symmetry is
\begin{equation}
\mathbf{A}_{2}= 2T\partial_{T} + X\partial_{X} + Y\partial_{Y}, \enskip %
f_{2}=0  \label{eq.PNerm2a}
\end{equation}
and the associated Noether FI
\begin{equation}
I_{2}= 2TH -(XX^{\prime }+YY^{\prime }).  \label{eq.PNerm2b}
\end{equation}

It can be shown that the three FIs $I_{0}, H, I_{2}$ are independent; therefore, the conservative generalized Ermakov system is superintegrable.

Although the remaining FIs will be expressible in terms of the $I_{0},H,I_{2}$, we continue in order to show that we recover the results of \cite{Leach 1994} which were obtained using Lie symmetries.

Case 3. The point Noether symmetry is
\begin{equation}
\mathbf{A}_{3} = T^{2}\partial_{T} + TX\partial_{X} +TY\partial_{Y}, \enskip %
f_{3}= \frac{X^{2}+Y^{2}}{2}  \label{eq.PNerm3b}
\end{equation}
with associated Noether integral
\begin{equation}
I_{3}= T^{2}H - T(XX^{\prime }+YY^{\prime }) + \frac{X^{2}+Y^{2}}{2}= \frac{%
I_{2}^{2}+2I_{0}}{4H}.  \label{eq.PNerm3c}
\end{equation}
We observe that the Lie
symmetries (2.9a), (2.9b), (2.9c) found in \cite{Leach 1994} are the point
Noether symmetries $\mathbf{A}_{1}$, (\ref{eq.PNerm2a}), (\ref{eq.PNerm3b}). Concerning the remaining FIs of \cite{Leach 1994} we have: (4.11) $I^{\prime }=2I_{0}$,
(4.12) $J_{1}^{\prime}=2H$, (4.13) $J^{\prime}_{2}=I_{2}$ and (4.14) $%
J^{\prime}_{3}=2I_{3}$. Using these relations, equation (4.18) is equivalent
to the expression (\ref{eq.PNerm3c}).

\subsection{The gradient KV $b_{1}\partial_{X} +b_{2}\partial_{Y}$}

\label{sec.erm.grKV}

The potential becomes\footnote{%
This is a superintegrable potential of the form $F(b_{1}Y-b_{2}X)$ (see
section 7 in \cite{MitsTsam 2020}).} $V_{1}= \frac{k}{(b_{1}Y - b_{2}X)^{2}}$ where $k,
b_{1}, b_{2}$ are arbitrary constants.

Case 2. The Noether generator, the Noether function and the FI are
\begin{equation*}
\mathbf{A}_{21}= b_{1}\partial_{X} +b_{2}\partial_{Y}, \enskip f_{21}= 0, %
\enskip I_{21}= b_{1}X^{\prime }+b_{2}Y^{\prime }.
\end{equation*}

Case 3. The Noether generator, the Noether function and the FI are
\begin{equation*}
\mathbf{A}_{31}= Tb_{1}\partial_{X} +Tb_{2}\partial_{Y}, \enskip %
f_{31}=b_{1}X +b_{2}Y, \enskip I_{31}= b_{1} (-TX^{\prime }+X) +b_{2}
(-TY^{\prime }+Y).
\end{equation*}

In order to compare these results with the ones of \cite{Naz 2020} we use polar coordinates $X=r\cos
\theta $ and $Y=r\sin \theta $. We find:\newline
\begin{equation*}
V_{1}= \frac{k}{r^{2}(b_{1}\sin\theta -b_{2}\cos\theta)^{2}}
\end{equation*}%
\begin{equation*}
\mathbf{A}_{21}= \left(b_{1}\cos\theta +b_{2}\sin\theta\right) \partial_{r}
+ \frac{1}{r} \left(b_{2}\cos\theta -b_{1}\sin\theta \right)
\partial_{\theta}, \enskip f_{21}= 0
\end{equation*}
\begin{equation*}
\mathbf{A}_{31}= \left(b_{1}T\cos\theta +b_{2}T\sin\theta
\right)\partial_{r} + \frac{1}{r}\left(b_{2}T\cos\theta -b_{1}T\sin\theta
\right) \partial_{\theta}, \enskip f_{31}= r(b_{1}\cos\theta +
b_{2}\sin\theta)
\end{equation*}
and
\begin{eqnarray*}
I_{21}&=& b_{1} \left( r^{\prime }\cos\theta - r\theta^{\prime }\sin\theta
\right) + b_{2} \left( r^{\prime }\sin\theta + r\theta^{\prime }\cos\theta
\right) \\
&=& b_{1} \left( \bar{p}_{1}\cos\theta -\frac{\bar{p}_{2}\sin\theta}{r}
\right) +b_{2} \left( \bar{p}_{1}\sin\theta +\frac{\bar{p}_{2}\cos\theta}{r}
\right)
\end{eqnarray*}
\begin{eqnarray*}
I_{31}&=& b_{1} \left(-Tr^{\prime }\cos\theta +Tr\theta^{\prime }\sin\theta
+r\cos\theta\right) +b_{2} \left(-Tr^{\prime }\sin\theta -Tr\theta^{\prime
}\cos\theta +r\sin\theta \right) \\
&=& b_{1} \left( -T\bar{p}_{1}\cos\theta + \frac{T\bar{p}_{2} \sin\theta}{r}
+ r\cos\theta \right) +b_{2}\left( -T\bar{p}_{1}\sin\theta - \frac{T\bar{p}%
_{2}\cos\theta}{r} + r\sin\theta\right)
\end{eqnarray*}
where $\bar{p}_{a}= \bar{\gamma}_{ab}\bar{q}^{b\prime}$ are the generalized
momenta. Replacing with $\bar{q}_{a}= (r,\theta)$ and $\bar{\gamma}%
_{ab}=diag(1,r^{2})$ we find that $\bar{p}_{1}=r^{\prime }$ and $\bar{p}%
_{2}= r^{2}\theta^{\prime }$.

It is straightforward to show that the point Noether symmetries $\mathbf{A}_{21}, \mathbf{A}_{31}$, the Noether
functions $f_{21}, f_{31}$ and the FIs $I_{21}, I_{31}$ are the symmetries $%
\mathbf{X}_{9}, \mathbf{X}_{10}$, the functions $B_{9}, B_{10}$ and the FIs $I_{9}, I_{10}$ respectively of \cite{Naz 2020}; while \\
- for $b_{1}=0, b_{2}=1$ they reduce to
the symmetries $\mathbf{X}_{5}, \mathbf{X}_{6}$, the functions $B_{5}, B_{6}$
and the FIs $I_{5}, I_{6}$ respectively of \cite{Naz 2020} and \\
- for $b_{1}=1$, $b_{2}=0$ they reduce to
the symmetries $\mathbf{X}_{7}, \mathbf{X}_{8}$, the functions $B_{7}, B_{8}$
and the FIs $I_{7}, I_{8}$ respectively of \cite{Naz 2020}.

As expected, the non-gradient KV (rotation) $Y\partial _{X}-X\partial _{Y}$ leads to the  linear first integral of angular momentum.
\bigskip

Finally, using the three FIs $H, I_{0}, I_{2}$, we  integrate the
system (\ref{eq.ermint8a}) - (\ref{eq.ermint8b}) and find that in polar coordinates the solution is
\begin{eqnarray}
r^{2}(T)&=& \frac{1}{2H}\left( 2HT-I_{2}\right) ^{2}+\frac{I_{0}}{H}  \label{eq.int2} \\
\int \frac{d\theta }{\sqrt{I_{0}-\bar{F}(\theta )}} &=& \pm \int \frac{\sqrt{2}}{%
r^{2}(T)}dT=\pm \frac{1}{\sqrt{I_{0}}}\tan ^{-1}\left[ \frac{1}{\sqrt{2I_{0}}%
}(2HT-I_{2})\right]  \label{eq.int3}
\end{eqnarray}%
where $\bar{F}(\theta )=(\tan ^{2}\theta +1)N(\tan \theta)$. In Table 3 of \cite{Naz 2020} the corresponding formula of (\ref{eq.int3}) gives $\tanh^{-1}$ instead of $%
\tan ^{-1}=\arctan $ which is the correct result.

\section{Conclusions}

Using recent results on the integrability of 2d conservative dynamical systems, we proved that the generalized Ermakov system is superintegrable and determined all the quadratic FIs. We showed that the recent results of \cite{Naz 2020} can be obtained from the more general method outlined in \cite{MitsTsam 2020} and \cite{Tsamparlis 2011}. Obviously, the methods discussed in the present work can be used by other authors in the study of similar dynamical systems. As a final remark, we note that an extension of this method, where one computes the time-dependent and autonomous  quadratic FIs without the use of Noether symmetries, has appeared recently in \cite{Mathematics 2021}.

\section*{Data Availability}

All data that supports the findings of this study are available within the article.

\section*{Conflict of interest}

The authors declare no conflict of interest.

\end{document}